\documentclass[aps,pre,amssymb,amsmath,twocolumn,floatfix]{revtex4-2}
\usepackage{graphicx}
\usepackage{dcolumn}
\usepackage{bm}

\graphicspath{ {./img/} }
\begin{document}

\title{Critical and geometric properties of magnetic polymers across the globule-coil transition}

\author{Kamilla Faizullina}
\author{Ilya Pchelintsev}
\author{Evgeni Burovski}
\affiliation{HSE University, 101000 Moscow, Russia}

\begin{abstract}
We study a lattice model of a single magnetic polymer chain, where Ising spins are located
on the sites of a lattice self-avoiding walk in $d=2$. We consider the regime where
both conformations and magnetic degrees of freedom are dynamic, thus the Ising
model is defined on a dynamic lattice and conformations generate an annealed
disorder. Using Monte Carlo simulations, we characterize the globule-coil and
ferromaget-to-paramagnet transitions, which occur simultaneously at a critical
value of the spin-spin coupling. We argue that the
transition is continuous---in contrast to $d=3$ where it is first-order.
Our results suggest that at the transition the metric exponent takes the
theta-polymer value $\nu=4/7$ but the crossover exponent  $\phi \approx 0.7$,
which differs from the expected value for a $\theta$-polymer.
\end{abstract}

\maketitle

\section{\label{sec:level1}Introduction}

A linear polymer in thermal equilibrium in a solvent can be either extended (``swollen''),
or collapsed into a dense globule, depending on the interplay between the excluded volume
effects, van der Waals attraction between monomers and its screening by the solvent \cite{deGennes1979}.
The physics of the phase transition between these two states, the so-called globule-coil transition
or $\theta$-transition,
is well captured by a simple lattice model of an interacting self-avoiding walk (ISAW),
with an attractive interaction between monomers on the nearest neighboring sites 
of the lattice \cite{Vanderzande1998}.

For magnetic polymers, where monomers carry magnetic moments (``spins''), 
the key parameter is the ratio of the relaxation times
of magnetic and conformational degrees of freedom \cite{Aerstens1992}: if spins
are fast, conformations generate a quenched disorder for the magnetic subsystem 
\cite{Aerstens1992, Chakrabarti1983, Chakrabarti1985, Papale2018};
in the opposite limit, the chain with quenched spins is qualitatively equivalent
to a disordered copolymer; several models of this kind have been discussed in
the literature \cite{Shakhnovich1994, KHOKHLOV1998, KHOKHLOV2012, Blavatska2014}.

The regime where both spins and conformations have comparable relaxation times
has so far received much less attention. In this regime, spins are defined
on a dynamic lattice, whose thermal fluctuations need to be taken into account
self-consistently, on an equal footing with spin fluctuations.
In this direction, Ref.\ \cite{Garel1999} introduced a model where monomers
of a SAW carry Ising spins, which interact via a short-range ferromagnetic
interaction. The model is investigated on a three-dimensional (3D) cubic lattice
using a mean-field approximation and Monte-Carlo (MC) simulations. In the absence of
external magnetic field, Ref.\ \cite{Garel1999} finds a first-order magnetic induced 
collapse transition---from a swollen paramagnetic phase to a ferromagnetic
globular phase. (Upon increasing the magnetic field, the transition is reported
to become continuous.)
In Ref.\ \cite{Faizullina2021} we considered a dynamic Hydrophobic-polar (HP)
model in two dimensions (2D). The collapse transition was found to be consistent
with a (continuous) $\theta$-transition of a nonmagnetic ISAW.

In this paper, we consider a ferromagnetic Ising model with spins placed on
a self-avoiding walk (SAW) on a 2D square lattice. Using MC simulations, we also
find a joint ferromagnetic and globule-coil transition, however our results
indicate that it is continuous---unlike the 3D model, where it is first order
\cite{Garel1999}. We argue that the transition is characterized by the
theta-point metric exponent $\nu$, but the crossover exponent $\theta$ is
markedly different. We also explore geometric properties of the model,
and stress the role of the surface terms.

\section{Model and method}

\begin{figure}[tbh]
\includegraphics[width=\columnwidth, keepaspectratio=True]{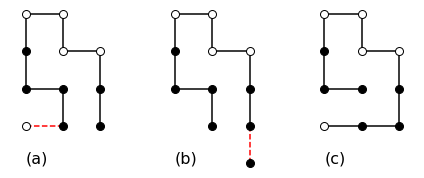} 
\caption{\label{fig:updates} Spin/SAW configurations and MC updates.
Straight lines show a sample SAW, open circles denote spins-up, $s_j=+1$, and
closed circles denote spins-down, $s_j=-1$. The BEE move is changing (a) to (b),
where the edge shown in dashed red line in (a) is removed and the edge shown in 
dashed red line in (b) is added. The reconnect update is changing the configuration between (a) and (c).
Note that configurations (a) and (c) have the same energy Eq.\ \eqref{hamiltonian}.}
\end{figure}

We consider the model of Ref.\ \cite{Garel1999}: 
Let $\mathcal{U}_N$ be a set of all SAW conformations of 
$N$ monomers joined by $N-1$ links on a 2D square lattice. Each monomer $i$ in a conformation
$u \in \mathcal{U}_N$ carries an Ising spin, $s_i = \pm 1$, see Fig.\ \ref{fig:updates}.
The spin-spin interaction
is short-ranged: two spins interact if they are nearest neighbors on the lattice.
Given a SAW conformation $u \in \mathcal{U}_N$ and
a sequence of $N$ spins, $\{s\}$, the Hamiltonian is
\begin{equation}
\label{hamiltonian}
E(\{s\}, u) = -J \sum_{ \langle i, j \rangle \in u} s_i s_j  - h\sum_{j \in u} s_j \;.
\end{equation}
Here the summation in the first term runs over pairs of spins, $i, j 
\in u$, which are nearest neighbors on the 2D lattice,
and $J > 0$ is the ferromagnetic exchange coupling. In the second term, 
$h$ is the magnetic field.

The partition function corresponding to Eq.\ \eqref{hamiltonian} reads
\begin{equation}
Z =\sum_{u \in \mathcal{U}_N} \sum_{\{s\}} e ^{-\beta E(\{s\}, u) },
\label{Zcanonical}
\end{equation}
where $\beta = 1/kT$ is the inverse temperature. To set the energy units, we
take $\beta = 1$ without loss of generality. Note that the summations in
Eq.\ \eqref{Zcanonical} run over both conformations and spin configurations.

For $h=J=0$, spins decouple from conformations, and the model
\eqref{hamiltonian}-\eqref{Zcanonical} reduces to a non-interacting SAW. In the
limit $h \gg J$, all spins are aligned, and Eq.\ \eqref{hamiltonian}-\eqref{Zcanonical}
reduces to the ISAW model. 
In this work we only consider the case $h=0$. 
In the limit $J \ll 1$, the model \eqref{hamiltonian}--\eqref{Zcanonical}
describes Ising spins located on a non-interacting SAW---for the spins, the
geometry is effectively one-dimensional and spontaneous magnetization is absent
in the thermodynamic limit \cite{Aerstens1992, Chakrabarti1983, Chakrabarti1985}. For $J \gg 1$, it is
natural to expect a dense ferromagnetically ordered globule.

We note that since Eq.\ \eqref{hamiltonian} only involves a single coupling constant,
it is natural to expect that the ferromagnetic ordering sets in simultaneously
with the globule-coil transition. In the next sections we verify this expectation
and characterize the corresponding transition.

\textit{Method.---} Most popular methods for Monte Carlo (MC) simulations of SAW-like model are
based on chain growth techniques with pruning and enrichment \cite{PermGrassberger},
and their flat-histogram generalizations \cite{Prellberg2004}.
We use a different strategy: we work directly with fixed-length configurations
and employ a variant of the worm algorithm \cite{Worm} for interacting SAW-like models \cite{BPS}.
Specifically, the method uses two sets of MC updates. First is a bilocal reptation update,
where we simultaneously remove a monomer from one end of a chain and add a monomer to the
other end--- the direction of the new edge and the value of the new spin are selected
at random, see Fig.\ \ref{fig:updates}(a)-(b).
This is nothing but the BEE move of Ref.\ \cite{Caracciolo2002}.
Second, to render the reptation dynamics ergodic and improve convergence 
for dense configurations, we also use the ``reconnect'' update, where we rotate a single
edge in the middle of the chain and attach it to the end of the chain---which needs
to be adjacent to an internal monomer, see Fig.\ \ref{fig:updates}(a)-(c).
The reconnect update is non-local since it reverses directions of $O(N)$ links of the SAW.
However the Metropolis acceptance probability \cite{Metropolis} equals unity since the update does
not change the energy, Eq.\ \eqref{hamiltonian}.
The reconnect update allows the simulation to escape from conformations where the end of
the chain is trapped inside a dense configuration \cite{BPS}.
Furthermore, to improve convergence of magnetic observables, we also use standard
Wolff cluster updates \cite{Wolff} for spins which keep the conformation fixed.

\section{Numerical simulations} 

We simulate our model on a square 2D
lattice for chains of up to $N = 10^4$ monomers. We typically use up to
$10^9$ MC updates for thermalization and collect statistics for
$10^{10}$ to $10^{11}$ MC steps. Here in a single MC step we select an update
(a BEE move, a reconnect or a spin cluster update) at random.

We perform simulations for $h=0$ and $0 < J < 2$. We collect
statistics for the mean energy, Eq.\ \eqref{hamiltonian}, per spin,
$\epsilon = \langle E \rangle / N$, the mean magnetization per spin,
$\langle m \rangle \equiv \langle \sum_{j \in u} s_j \rangle / N$ and its powers,
$\langle m^2 \rangle$ and $\langle m^4 \rangle$. To characterize the structural
properties of the model, we measure the mean end-to-end distance of the SAW,
$\langle R^2_N \rangle$.
\footnote{In the literature, the gyration radius is often considered instead;
however the asymptotic properties of the radius of gyration and end-to-end
distance are expected to be the same (see, e.g. \cite{Caracciolo2002}), and the
latter is simpler to work with numerically.
}
Here and elsewhere in the text, angular brackets denote the MC average approximating
the average over the Gibbs distribution \eqref{Zcanonical}.

\begin{figure}[h]
	\includegraphics[width=\columnwidth, keepaspectratio=True]{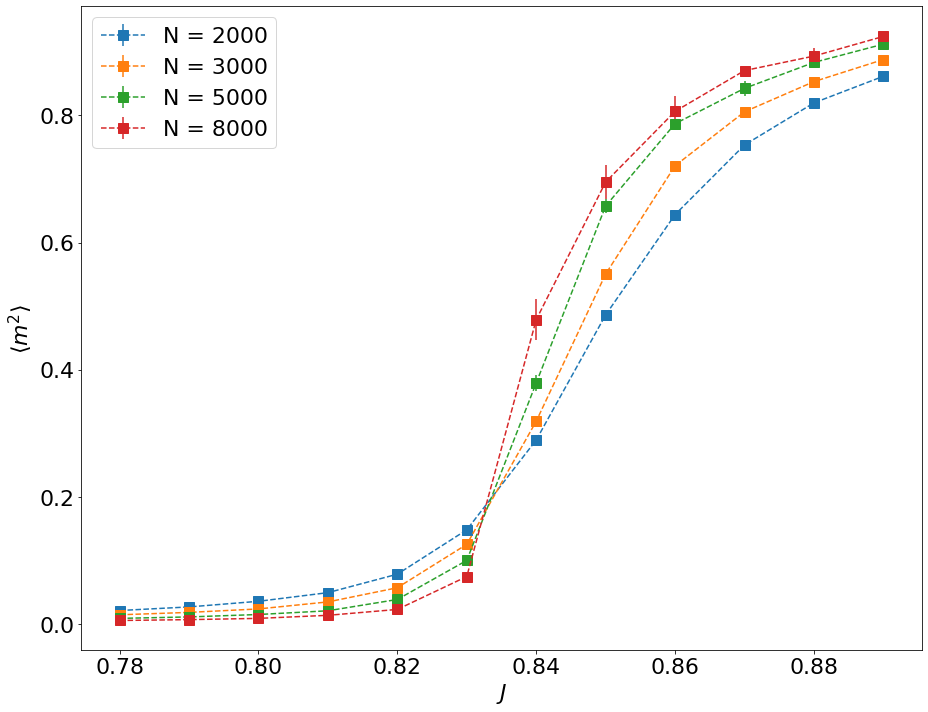} \\%
	\includegraphics[width=\columnwidth, keepaspectratio=True]{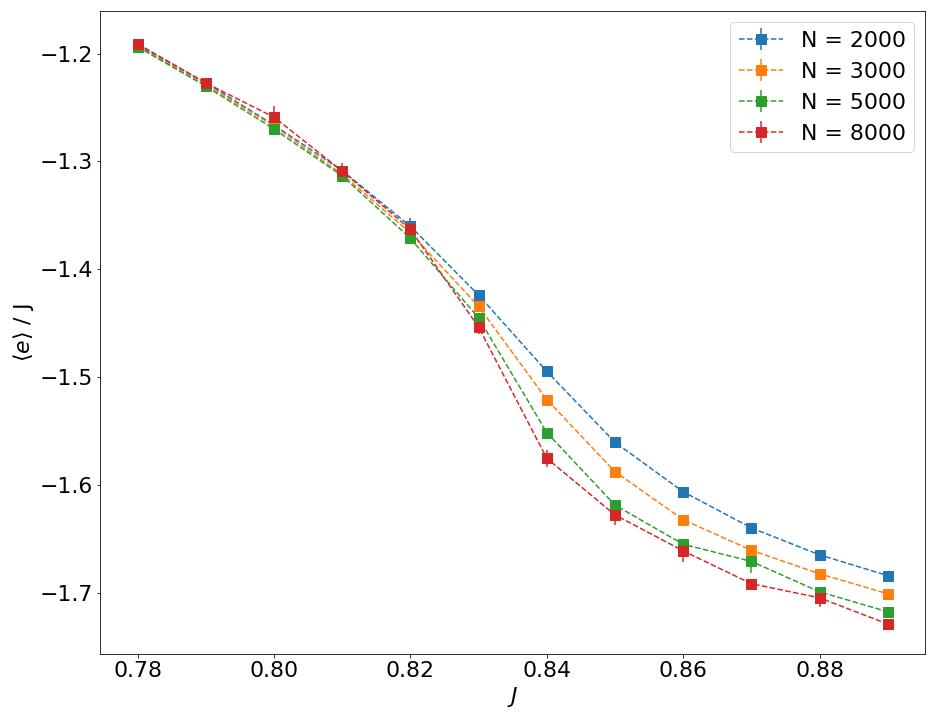} 
	\caption{\label{fig:2dIsing} (top) mean squared magnetization as a function
		of $J$ for several values of $N$. Solid squares with errorbars are MC results, lines are
		to guide an eye only. Errorbars are estimated via binning analysis. In these simulations we use at
		least $7 \times 10^{9} $ MC steps per data point.
		(bottom) Mean energy as a function of $J$ for several values of $N$. Squares are MC data with
        errorbars, and lines are to guide an eye. See text for discussion.
	}
\end{figure}

Fig.\ \ref{fig:2dIsing}(top) shows simulation results for mean square 
magnetization, $\langle m^2 \rangle$, as a function of $J$ for several
representative values of the SAW lengths $N$. At small values of $J$, $\langle m^2 \rangle \to 0$ at
increasing $N$, which is consistent with the spontaneous magnetization being zero in the
thermodynamic limit \cite{Aerstens1992, Chakrabarti1983, Chakrabarti1985}.
For larger values of the coupling constant, magnetization increases with
increasing $J$ and starts saturating for $J \gtrsim 0.88$,
which suggests a ferromagnetic ordering for large $J$.

Fig.\ \ref{fig:2dIsing}(bottom) illustrates the behavior of the mean energy,
which approaches the asymptotic $N\to\infty$ value of $-2J$ for a densely packed
fully magnetized walk. Finite-size corrections are clearly visible for both
$\langle m^2 \rangle$ and $\langle \epsilon \rangle$, and we note that corrections
are more pronounced for $J \gtrsim 0.82$, especially in Fig.\ \ref{fig:2dIsing}(bottom).

\begin{figure}[h]
	\includegraphics[width=0.99\columnwidth, keepaspectratio=True]{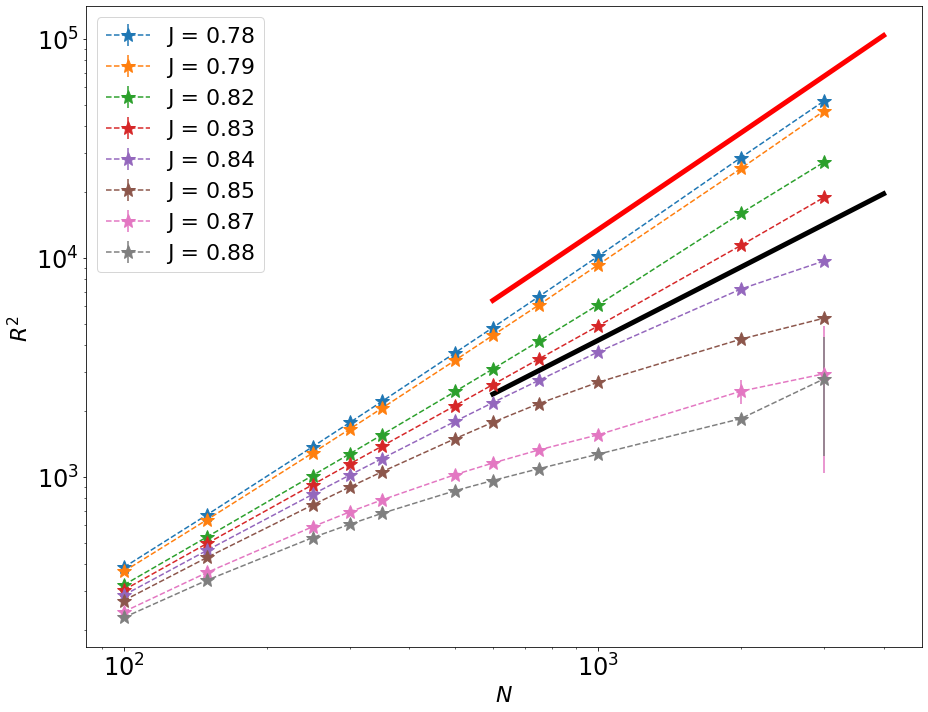} \\%
	\caption{\label{fig:R2_logplot}
		Mean squared end-to-end distance as a function of $N$ from $N=100$ to
        $N=3000$ for several values of $J$. Stars are MC data with errorbars,
        dashed lines are to guide an eye, and solid lines are $R^2 \sim N^{2\nu}$
        with $\nu = 3/4$ (the solid red line) and $\nu = 4/7$ (the solid black line).
        See text for discussion.
	}
\end{figure}

Fig.\ \ref{fig:R2_logplot} shows the dependence of the mean end-to-end distance,
$\langle R^2_N \rangle$, on $N$ for several values of the coupling constant $J$.
For $N \gg 1$ the scaling is visually consistent with a power-law,
\begin{equation}
\langle R^2_N \rangle \sim N^{2\nu} (1 + \cdots)\;,
\label{R2nu}
\end{equation}
where dots represent corrections-to-scaling. For comparison,
Fig.\ \ref{fig:R2_logplot} also shows the asymptotic power laws $N^{2\nu}$ with
$\nu=3/4$---which is a non-interacting SAW value (see e.g.,\cite{Rensburg2015}),---
and $\nu = 4/7$---which is the exact value for the 2D ISAW at the
$\theta$-point \cite{Duplantier1987}.

Numerical data in Fig.\ \ref{fig:R2_logplot} seem to indicate that the scaling
of the end-to-end distance for our model crosses over from a non-interacting SAW
limit for small $J$ to a $\theta$-point scaling for $J \sim 0.83$, and further
on towards $\nu = 1/2$, which is expected for a dense globular phase.
\footnote{
Following Ref.\ \cite{Berretti1985}, we also fit the data shown in
Fig.\ \ref{fig:R2_logplot} with a four-parameter model,
$\log (R_N^2+k_1 ) = 2 \nu \log (N+k_2) + b$, with fit parameters $\nu$, $b$, $k_1$ and $k_2$.
Here $b$, $k_1$ and $k_2$ are phenomenological parameters
meant to mimic corrections to scaling. While this model is not expected to be fully
accurate---it misrepresents corrections-to-scaling exponents 
and thus produces wrong results for $\nu$ close to the globular phase---it does
support the expectation from a visual inspection of Fig.\ \ref{fig:R2_logplot} that the metric
exponent $\nu$ agrees with the $\theta$-point value $\nu=4/7$ around $J\approx 0.83$.
} 
Taken together, our numerical results shown in Figs.\ \ref{fig:2dIsing} and
\ref{fig:R2_logplot}, indicate that both magnetic and structural properties of the model
undergo a change at around $J\sim 0.83$.

\begin{figure}[hbt]
\includegraphics[width=\columnwidth, keepaspectratio=True]{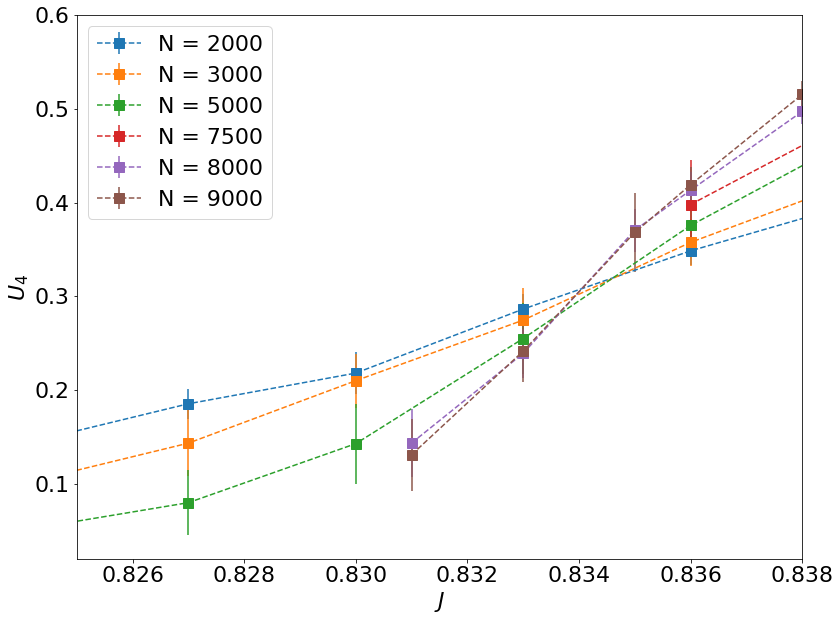} \\%
\includegraphics[width=\columnwidth, keepaspectratio=True]{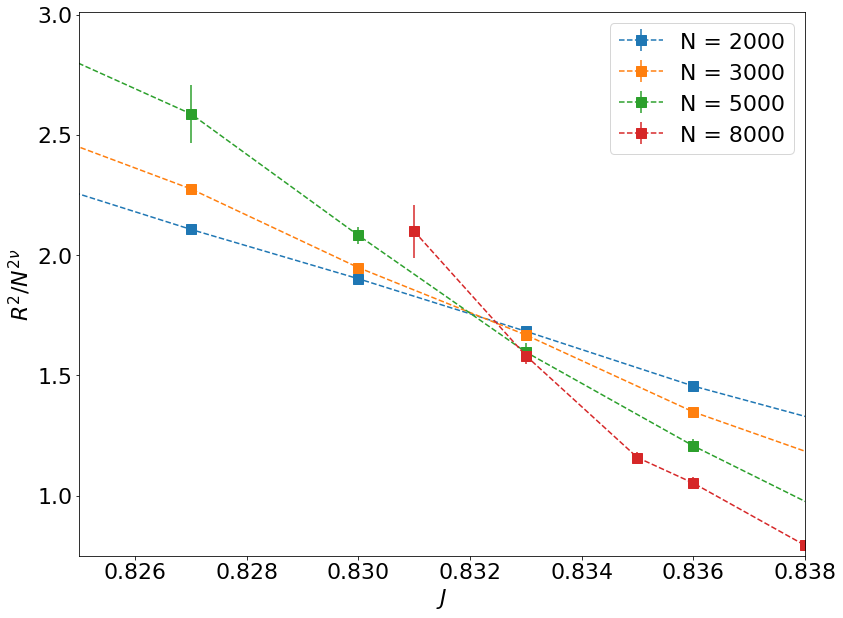} 
\caption{\label{fig:U4_R2} (top) Binder cumulants \eqref{cumulant} as a function
of $J$ for several values of $N$. Solid squares with errorbars are MC results, lines are
to guide an eye only. Errorbars are estimated via a Gaussian resampling from errorbars
of $\langle m^4\rangle$ and $\langle m^2 \rangle$. 
%
(bottom) Scaled mean end-to-end distance \eqref{R2nu} with $\nu = 4/7$,
which is the exact value for the 2D ISAW at the $\theta$-point \cite{Duplantier1987}.
Squares are MC data with errorbars, and lines are to guide
an eye. See text for discussion.
}
\end{figure}

\textit{The joint transition.---}%
To locate the magnetic transition between paramagnetic and ferromagnetic phases,
we compute the fourth-order Binder cumulant, 
\begin{equation}
U_4 = 1 - \frac{ \langle m^4 \rangle}{3 \langle m^2 \rangle^2} \;,
\label{cumulant}
\end{equation}
which is expected to become scale-independent at the transition \cite{Binder1981}.

Fig.\ \ref{fig:U4_R2}(top) shows the dependence of the Binder cumulant
\eqref{cumulant} on interaction $J$ for several values of $N$. For large values of
the coupling constant (not shown in Fig.\ \ref{fig:U4_R2}), $U_4$ tends to the
value $2/3$ from below, as expected for a ferromagnetic
state \cite{Binder1981}. Curves of the cumulant $U_4$ for varying $N$ 
cross around $J \approx 0.834$, indicative of the paramagnetic-to-ferromagnetic phase
transition. Finite-size corrections are clearly visible in Fig.\ \ref{fig:U4_R2}(top),
thus to get a more precise estimate for the transition temperature, we analyze 
the pairwise crossings of the $U_4$ vs $N$ curves for a series of $N$ values
from $N=2000$ to $N=9000$. The final estimate for the critical values is
\begin{equation}
J_c = 0.8340(5)\,,\qquad U_4^{(c)} = 0.308(8)\;.
\label{Jcrit}
\end{equation}
This result \eqref{Jcrit} is close to, but distinct from the estimate
$J_c = 1/1.18 \approx 0.847$, stated as preliminary without much discussion in
Ref.\ \cite{Garel1999}.

Fig.\ \ref{fig:U4_R2}(bottom) shows the dependence of the mean squared 
end-to-end distance \eqref{R2nu}. Here we rescale the values of $R^2_N$ by
$N^{2\nu}$ with $\nu = 4/7$, as suggested by the analysis in the previous section.
With this rescaling, $\langle R^2\rangle/N^{2\nu}$ becomes $N$-independent
(modulo corrections-to-scaling) at $J_\theta = 0.833(1)$ which is consistent with
Eq.\ \eqref{Jcrit} within the combined errorbars.

We also checked that the existence of the crossing is sensitive to the value
of the metric exponent $\nu$: if $\nu$ is changed by more then 0.07,
the crossing disappears.

We thus conclude that our numerical data suggest that (i) the ferromagnetic
and globule-coil transition occur simultaneously at the critical coupling
constant given by Eq.\ \eqref{Jcrit}, and (ii) the scaling of the end-to-end
distance at the transition is consistent with the $\theta$-point metric exponent $\nu=4/7$.

\begin{figure}[hbt]
\includegraphics[width=0.99\columnwidth, keepaspectratio=True]{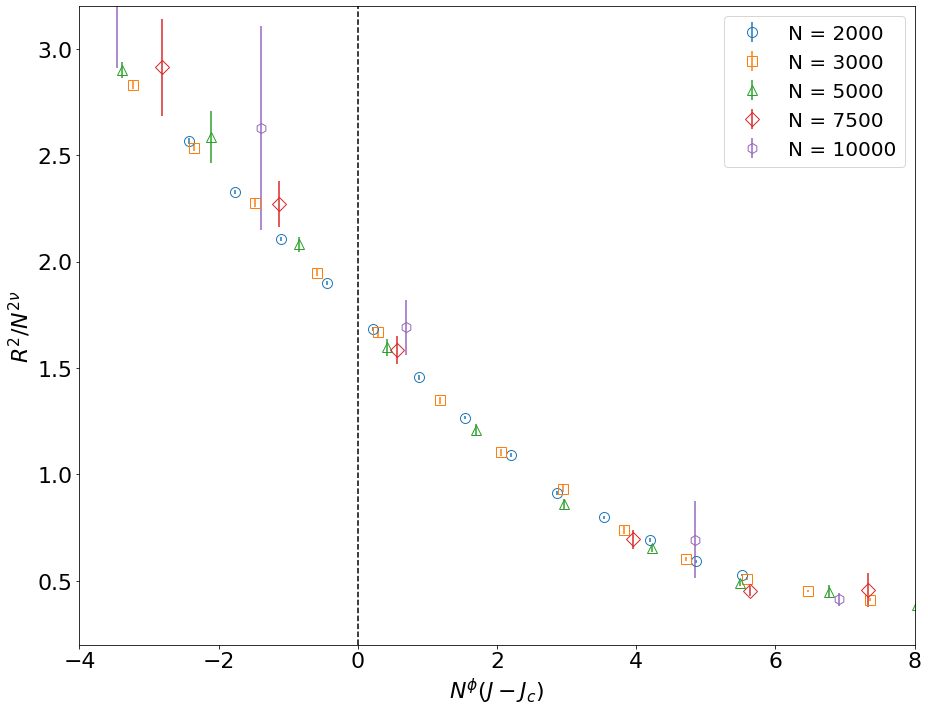} \\%
\caption{\label{fig:R2_phi} Data collapse for the scaled end-to-end distance,
$\langle R^2_N \rangle / N^{2\nu}$, vs the scaled coupling
$x = (J - J_c) N^\phi$. We fix $\nu=4/7$ and vary $J_c$ and $\phi$.
On this plot, $J_c=0.832$ and $\phi=0.7$. From visual inspection of the quality of the 
collapse, we estimate $J_c = 0.833(1)$ and $\phi = 0.7(1)$. See text for discussion.
}
\end{figure}

\textit{The crossover exponent.---}%
We turn our attention to estimating the crossover exponent $\phi$ which
quantifies the deviation from criticality via the scaled coupling $x = (J-J_c) / N^{-\phi}$
 \cite{Rensburg2015}. Specifically, the end-to-end distance is expected to follow
$\langle R^2_N\rangle  = N^{2\nu} f(x)$ where $f(\cdot)$ is a dimensionless
function of a dimensionless variable. To probe this Ansatz, we perform data collapse
of the end-to-end distance, where we keep $\nu=4/7$ fixed at its theta-point value \cite{Duplantier1987},
and vary $J_c$ and $\phi$. This procedure is illustrated in Fig.\ \ref{fig:R2_phi}.

We find that our MC data are consistent with $J_c = 0.833(1)$ and $\phi=0.7(1)$,
where the errorbars are conservative estimates from visual inspection of the
quality of the data collapse. We note that the value of $J_c$ is consistent with
Eq.\ \eqref{Jcrit}. The crossover exponent clearly differs from the $\theta$-point
value for the ISAW model, where the Coulomb gas prediction is
$\phi=3/7$ \cite{Duplantier1987} and numerical estimates are somewhat
larger (see Ref.\ \cite{Caracciolo2011} and 
the discussion therein). 

\begin{figure}[hbt]
	\includegraphics[width=0.99\columnwidth, keepaspectratio=True]{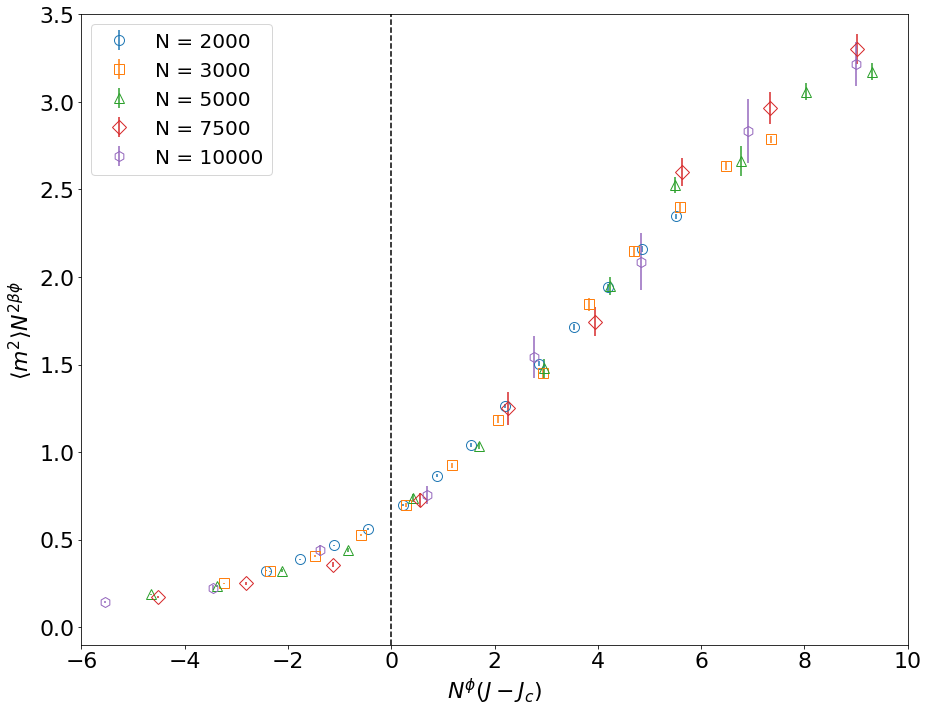} \\%
	\caption{\label{fig:mag_scale} Data collapse for the second moment of magnetization $\langle m^2 \rangle$.
    In this plot we use $\phi = 0.71$, $J_c = 0.832$ and $\beta = 1/8$.
    See text for discussion.
	}
\end{figure}

We also perform a similar data collapse analysis for the magnetization, where 
the scaling Ansatz is $\langle m^2 \rangle = N^{-2\beta\phi} g(x)$, where $g(x)$
is a scaling function and $\beta$ is the order parameter exponent.
Fig.\ \ref{fig:mag_scale} illustrates the procedure where we take
$\beta=1/8$---which is the value for the 2D Ising universality class. While the
quality of our numerical data does not allow for estimating critical exponents
with accuracy of any less then, say, 50\%, we find that our data are consistent with the order
parameter exponent taking the 2D Ising value, and the crossover exponent $\phi \approx 0.7$.

We stipulate that a high-precision estimate of the crossover exponent and/or the
order parameter exponent should
take into account two sources of corrections. First, for a disordered Ising model,
logarithmic corrections \cite{Dotsenko1983}, are known to lead to apparently
varying exponents \cite{Ballesteros1997}. Second, non-universal corrections
due to the surface tension are strong for 2D SAWs \cite{Grassberger1995}
because the surface-to-volume ratio in 2D scales as $\sim N^{-1/2}$ which is
close to the universal $\theta$-point values $\nu=4/7$ and $\phi=3/7$.

\textit{Bulk to surface ratio.---} Strictly speaking, the very notions of bulk
and surface are not well defined for $J < J_c$, where typical conformations
are coil-like. To come up with a quantitative characteristic which is meaningful
across the globule-coil transition and can be interpreted as a bulk-to-surface
ratio in the globular phase, we consider a local neighborhood of 
a monomer. We note that each monomer (apart from two endpoints of the chain) can
be classified according to the number of its neighbor monomers as being either
1D-like (two neighbors), 2D-like (four neighbors) or surface-like (3 neighbors).

For a length-$N$ conformation, we count the numbers of monomers of each kind; 
dividing by $N$ we obtain the fractions, $n_\alpha$ ($\alpha = 2, 3, 4$), 
so that $n_2 + n_3 + n_4 = 1 - 2/N$. Qualitatively, the ratio $n_2 / (n_3 + n_4)$
characterizes a blob-and-link structure of a coil-type conformation, and 
$n_4/n_3$ can be interpreted as a proxy for a bulk-to-surface ratio.

\begin{figure*} 
\includegraphics[keepaspectratio=True, width=0.99\textwidth]{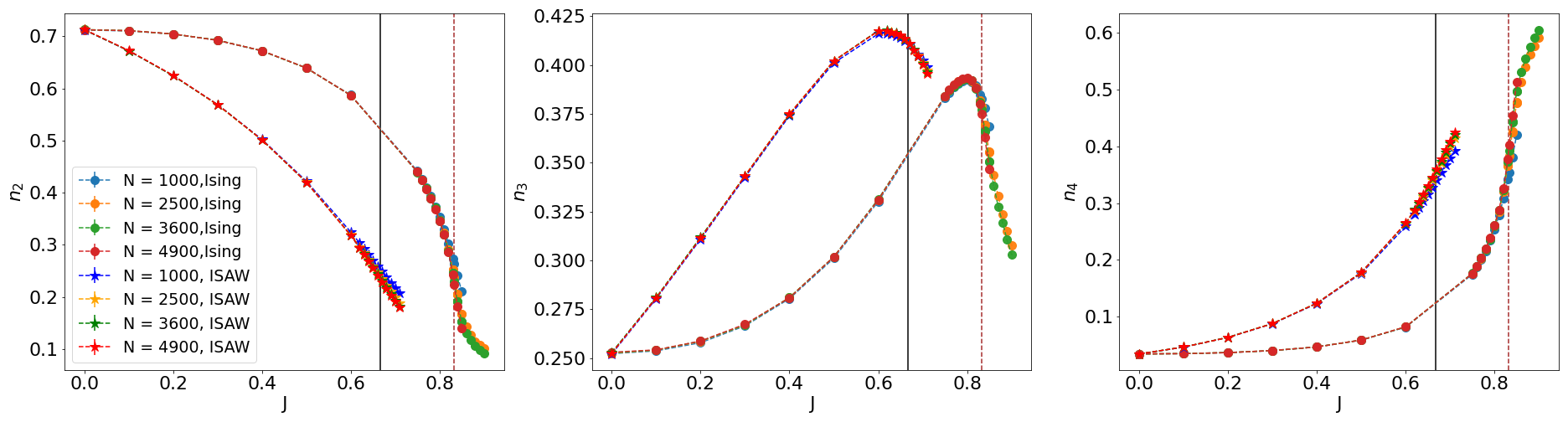} 
\caption{\label{fig:bulk} Fractions of monomers with two neighbors, $n_2$, (left),
three neighbors, $n_3$, (center) and four neighbors, $n_4$, (right).
Solid circles are the MC data for the Ising model \eqref{hamiltonian}-\eqref{Zcanonical}, 
stars are the MC data for the ISAW model, and dotted lines are to guide an eye.  
The vertical solid black line is the theta-point for the ISAW, taken from
Ref.\cite{Caracciolo2011}. The vertical dashed brown line is Eq.\ \eqref{Jcrit}. }
\end{figure*}

Fig. \ref{fig:bulk} shows the fractions of each kind of monomers as a function
of $J$  for chains of length $N=1000$ to $4900$. For comparison, we also compute
the corresponding fractions for an ISAW model (i.e., Eqs.\ 
\eqref{hamiltonian}-\eqref{Zcanonical} with $h \gg J$).

Several features stand out in Fig.\ \ref{fig:bulk}. First, even in the
non-interacting SAW limit, $J\to 0$, conformations are not fully 1D-like, as
$n_2 \approx 0.75$ only (the finite-size corrections become negligible for 
$N \gtrsim 100$). The ``bulk'' fraction, $n_4$, is vanishingly
small in the $J\to 0$ regime, and the fraction of the ``surface'' monomers, $n_3$,
tends to 0.25 for $J\to 0$. 
In the opposite limit of large $J$, the 1D-like fraction tends to zero and
the ``bulk'' fraction grows. Most surprisingly, the ``surface'' fraction, $n_3$,
develops a peak for both ISAW and Ising models in the vicinity of their
respective collapse transitions.

While the relation between these results to a bulk/surface ratio of real 
polymer chains is qualitative at best, and that more work is needed to understand
the nature of the peaks of $n_3(J)$, these results do illustrate the importance
of surface effects and stress the qualitative difference between the magnetic
SAW models and spin networks with mixed 1D / 2D local connectivity \cite{Dasgupta2014}.

\textit{Relation to the Ising model on rectangular lattices.---}  
 
It is instructive to compare the critical value of the Binder cumulant, $U_4^{(c)}$,
Eq.\ \eqref{Jcrit}, to the values for a usual Ising model 
on a regular grid. For the Ising model on a rectangular $L\times W$ lattice, 
the critical value of $U_4$ depends on the boundary conditions and on the aspect
ratio of the lattice, $L/W$ \cite{SelkeShchur2005, Selke2006}. The dependence on the boundary
conditions is strong: on an $L\times L$ lattice with periodic boundary conditions,
$U^{(c)} \approx 0.61$, while open boundary conditions lead to
$U^{(c)} \approx 0.4$. Furthermore, on the lattice with open boundary conditions, $U^{(c)}$ decreases continuously for
increasing aspect ratio $L/W$ down to $\approx 0.35$ for $L/W=2$ \cite{Selke2006}
and further down for larger aspect ratios.

The critical value $U_4^{(c)}$, Eq.\ \eqref{Jcrit}, is approximately compatible with
the result for the Ising model on a rectangular lattice with open boundary conditions
and the aspect ratio given by the ratio of the eigenvalues of the gyration tensor
of an interacting SAW at the $\theta$-point \cite{Caracciolo2011}. More work is
needed to accurately trace this connection.

\textit{The nature of the transition.---} In 3D, the transition is clearly
first-order \cite{Garel1999}. Our simulations indicate that the transition
is continuous in 2D.  First of all, the Binder cumulant
\eqref{cumulant} is a monotonic function of $J$ for fixed $N$,
cf Fig.\ \ref{fig:U4_R2}(top).
This is consistent with a continuous transition, and is in contrast to the 
expected behavior for a first-order transition, where the cumulant is
non-monotonic and develops a dip at $J_c$ as $N$ increases \cite{BinderLandau1984}. 

We then perform simulations for the specific heat capacity per monomer, which is
given by the second moment of the energy, Eq.\ \eqref{hamiltonian}--\eqref{Zcanonical}, 
\begin{equation}
C = \frac{1}{N} \left( \langle E^2 \rangle -  \langle E \rangle^2 \right)    
\label{heatcapacity}
\end{equation} 
For finite values of $N$, the heat capacity is expected to have a peak in the
critical region. The peak can be rounded and shifted by finite-size corrections,
and the evolution of the peak height and shape is expected to be very different
for first-order and continuous transitions: For a first order transition, the
height of the peak of $C(J)$ is expected to be linear in $N$, while the width
is expected to shrink as $\sim N^{-1}$ \cite{BinderLandau1984}. For continuous
transitions, the structure of $C(J)$ in the vicinity of $J_c$ is controlled by
the heat capacity exponent $\alpha$, which is typically different from unity.

\begin{figure}[htb]
	\includegraphics[width=0.99\columnwidth, keepaspectratio=True]{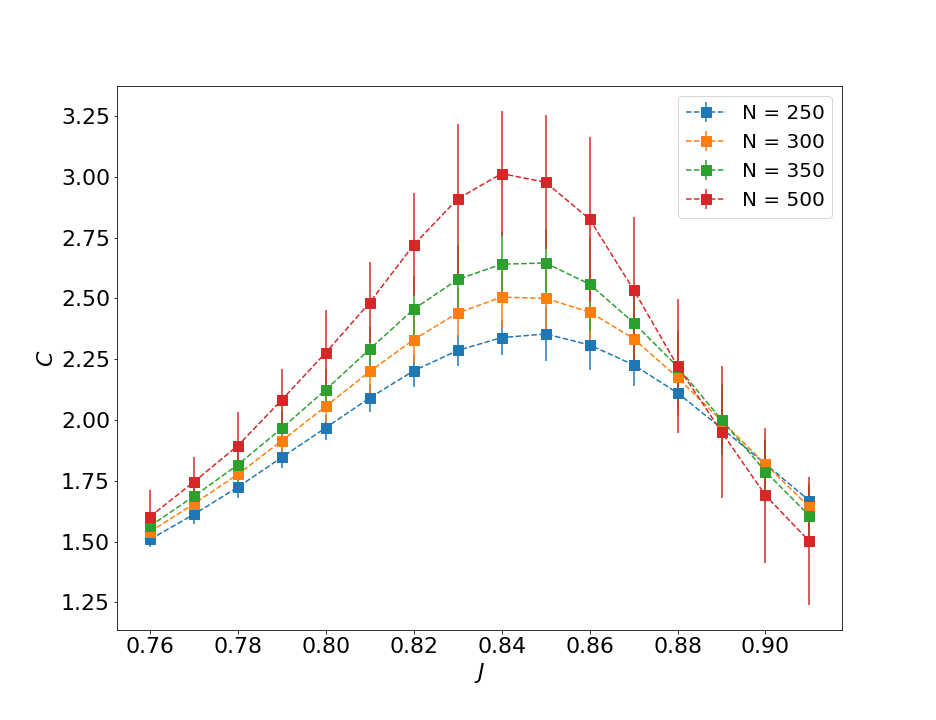} \\%
	\caption{\label{fig:heat}The specific heat capacity per monomer,
             Eq.\ \eqref{heatcapacity}, as a function of the coupling constant $J$.
             Errorbars are estimated via statistical resampling from MC data for
             the first and second moments of the energy. See text for discussion.
	}
\end{figure}

Fig.\ \ref{fig:heat} shows our numerical results for the specific heat capacity.
We note that numerical cancellations in Eq.\ \eqref{heatcapacity} magnify
statistical errors of MC simulations, thus limiting the values of $N$ accessible
in these simulations to be about  an order of magnitude smaller then
those in Figs.\ \ref{fig:2dIsing}-\ref{fig:mag_scale}---which is comparable
to the values reported in Ref.\ \cite{Garel1999}.
At these values of $N \leqslant 500$, shown in Fig.\ \ref{fig:heat},
finite-corrections are very strong.
Nevertheless, the available numerical data suggest that the peak height
dependence on $N$ is sublinear and the peak widths shrinks slower then $1/N$.
The overall shape of $C(J)$ curves in Fig.\ \ref{fig:heat} is drastically different
from those observed for a first-order transition in 3D in Ref.\ \cite{Garel1999}.
We interpret these observations, however limited, as an additional indication of
a transition being continuous, with the heat capacity exponent $\alpha < 1$.

We also note that we observe a single peak of $C(J)$, not a two-peak structure
reported for a site-diluted Ising model \cite{SelkeShchurVasiliev1998} and a
network of Ising spins with mixed 1D/2D local connectivity \cite{Dasgupta2014}.
The difference with the latter is not surprising given the role of the
surface-like spins, cf Fig.\ \ref{fig:bulk}.

\begin{figure}[htb]
\includegraphics[scale=0.26]{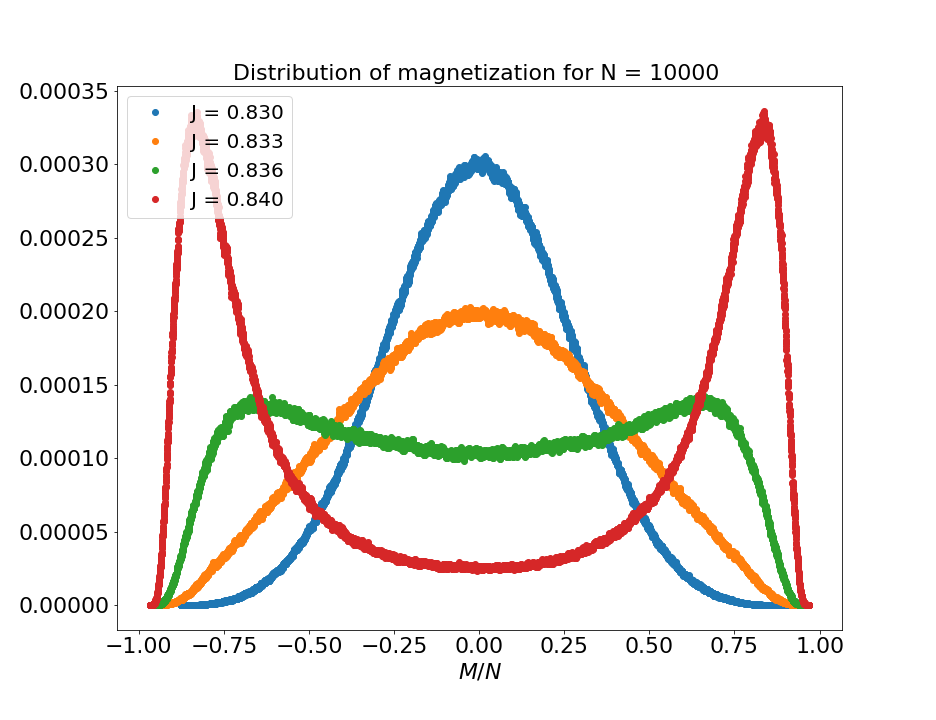} 
\caption{\label{fig:magnetization10000}  Distribution of the magnetization
$m = \sum_{j\in u} s_j$/N for $N=10000$. The coupling constants are
$J=0.830 < J_c$ (blue points), $J=0.833 \approx J_c$ (orange), $J=0.836$ (just
above the $J_c$, green), and $J = 0.840 > J_c$ (red). 
Each simulation uses $\sim 7\times 10^{9}$ MC steps.}
\end{figure}

To further check the nature of the transition, we compute
distributions of observables. Fig.\ \ref{fig:magnetization10000} shows the
distribution of the magnetization for $N=10^4$ in the vicinity of the transition,
Eq.\ \eqref{Jcrit}. The distribution is Gaussian-like
on the paramagnetic side, $J < J_c$, broadens on approach to the critical coupling,
and develops a clear ferromagnetic structure ($m = \pm 1$) for $J > J_c$. In the
critical region, we see no signs of a phase coexistence which would signal a
first-order transition.

\section{Conclusions and outlook}

\textit{Concluding,} we study a 2D model of a magnetic polymer chain where
monomers of a self-avoiding walk on a lattice carry Ising spins \cite{Garel1999}. 
We use a variant of the worm algorithm to simulate fixed-length chains of up to 
$10^4$ monomers. We find a joint transition---where both spins order
ferromagnetically and the SAW collapses into a globular phase---at
$J/T = 0.8340(5)$. The very fact that the transitions occur simultaneously can
be traced to the specifics of the model, which only has a single coupling
constant, the exchange integral for the short-range spin-spin interaction. 
What is less clear \textit{a priori}, is the nature of the transition. Our
results suggest that the transition is continuous, in contrast to a similar
3D model, where it is reported to be first-order \cite{Garel1999}.
Our numerical results suggest that some critical exponents (but not all of them)
are inherited from the ``parent'' models, namely the $\theta$-polymer ISAW model,
and the Ising model. 
Specifically, we present numerical evidence that the metric exponent $\nu$ at the
transition takes the $\theta$-point value $\nu=4/7$, but the crossover
exponent $\phi \approx 0.7$, which is clearly different from the $\theta$-polymer
value of $3/7$. 
We also present indications that the magnetic order parameter exponent $\beta$
is consistent with the 2D Ising universality class value $\beta=1/8$, however 
the accuracy of this observation given our simulation results is relatively weak. 

We study geometric properties of the model and classify the local connectivity
of monomers of the chain into 1D-like, bulk-like and surface-like. A possible
interpretation of our numerical results is that the surface-to-bulk ratio
has a peak in the vicinity of the transition. Incidentally, we also find 
numerical evidence that for a non-interacting SAW, the fraction of 1D-like monomers
is 1/4 in the thermodynamic limit. To the best of our knowledge, this was previously not
discussed in the literature. More work is needed to clarify the status and physical
meaning of these numerical results.

Concerning future work, it would be interesting to explore more
realistic models of magnetic polymers, e.g. by considering Potts or Heisenberg type
models and general dipole-dipole couplings in two and three dimensions.
Models with separate coupling constants might generate richer phase diagrams
with separate globule-coil and magnetic transitions.

Possible experimental realizations of magnetic polymers, for which our model and
its suggested generalizations may be applicable, include magnetic
filaments where magnetic nanoparticles are either cross-linked by a polymer to 
form linear structures---these can be realized via e.g. biotemplaing
\cite{BereczkTompa2017}--- or self-organize into one-dimensional like structures
at liquid-liquid interfaces \cite{Benkoski2008}. 
Monte Carlo simulations of models of magnetic polymers may complement molecular
dynamics studies of magnetic filaments \cite{Mostarac2020}.

When this work was completed, we became aware of an independent study
of the same model in Ref.\ \cite{Foster2021}. Our estimates of the location of
the transition and critical exponents and those of Ref.\ \cite{Foster2021}
are consistent within the combined errorbars.

\section{Acknowledgments } 
We acknowledge financial support by RFBR according to the research project
No 19-07-01117. K.F. and I.P. acknowledge support within the Project Teams
framework of MIEM HSE. Numerical simulations were performed using the  computational
resources of HPC facilities at HSE University \cite{Kostenetskiy_2021}.
Multiple illuminating discussions with Lev Shchur and Yury Budkov are gratefully
acknowledged.

\bibliography{bibliography}

\end{document}